\documentclass[namedreferences]{kluwer}

\usepackage[dvips]{graphicx}

\newbox\grsign \setbox\grsign=\hbox{$>$} 
\newdimen\grdimen \grdimen=\ht\grsign
\newbox\laxbox \newbox\gaxbox
\setbox\gaxbox=\hbox{\raise.5ex\hbox{$>$}\llap
     {\lower.5ex\hbox{$\sim$}}}\ht1=\grdimen\dp1=0pt
\setbox\laxbox=\hbox{\raise.5ex\hbox{$<$}\llap
     {\lower.5ex\hbox{$\sim$}}}\ht2=\grdimen\dp2=0pt
\def\gax{\mathrel{\copy\gaxbox}}
\def\lax{\mathrel{\copy\laxbox}}

\begin{document}

\begin{article}

\begin{opening}

\title{EUVE Observations of VW Hydri in Superoutburst}
\author{CHRISTOPHER W.\ MAUCHE}
\runningauthor{CHRISTOPHER W.\ MAUCHE}
\runningtitle{EUVE OBSERVATIONS OF VW HYDRI}
\institute{%
 Lawrence Livermore National Laboratory, \\
 L-41, P.O.~Box 808, Livermore, CA 94550}

\begin{abstract}
{\it EUVE\/} observed the SU~UMa-type dwarf nova VW~Hydri in superoutburst
for an interval of nearly 2 days in 1994 June and produced EUV light curves
and the first EUV spectrum of this important CV.
\end{abstract}

\end{opening}

\vbox to -12pt{\vskip -7.00cm
\hbox to \hsize{{\it 1996, Cataclysmic Variables and Related Objects,\hfil }}
\hbox to \hsize{{\it ed.~A.\ Evans \& J.\ H.\ Wood (Dordrecht: Kluwer),
243\hfil }}
\vss}
\vskip -12pt

\vskip -12pt
\section{Observations}

VW~Hyi was observed in superoutburst with {\it EUVE\/} from 1994 June 2.13 to
4.10 UT (RJD 9505.63 to 9507.60; $\rm RJD=JD-2440000$). The observations took
place $\approx 3.5$ days after the rise of the optical flux on $\approx $
May 29.5 UT and $\approx 2$ days after the peak of the optical flux at
$V\approx 8.5$ on $\approx $ May 31.0 UT. The Deep Survey photometer (70--170
\AA) collected data for less than one day before being shut off because of the
high count rate ($\sim 20$ counts~s$^{-1}$) induced in the detector by source
photons. Earth occultations and standard settings on the pipeline reduction
software presently limits the useful amount of spectrometer data to 4280, 4177,
and 1075~s for the SW (70--180~\AA ), MW (145--370~\AA ), and LW (290--760
\AA ) spectrometers, respectively.

As measured by the Deep Survey photometer and the spectrometers, the EUV flux
of VW~Hyi was highly variable during the observations. The upper panel of
Figure~1 shows the count rates recorded by the SW, MW, and LW spectrometers in
the wavebands 80--170, 170--300, and 300--400~\AA , respectively. Whereas the
visual flux {\it fell\/} steadily from $V\approx 8.7$ to $V\approx 8.9$ during
the observations, the EUV flux is seen to {\it increase\/}. On shorter time
scales, the EUV flux variations, in particular the flux diminutions, are
accompanied by distinct spectral variations. The lower panel of Figure~1 shows
the ratio of the count rates in the 80--170~\AA \ to 170--300 \AA \ wavebands
recorded by the SW and MW spectrometers as a function of time. Each diminution
of the EUV flux is accompanied by a softening of the spectrum; just the reverse
of the effect produced by variations in the column density. 

\begin{figure}
\centerline{\includegraphics{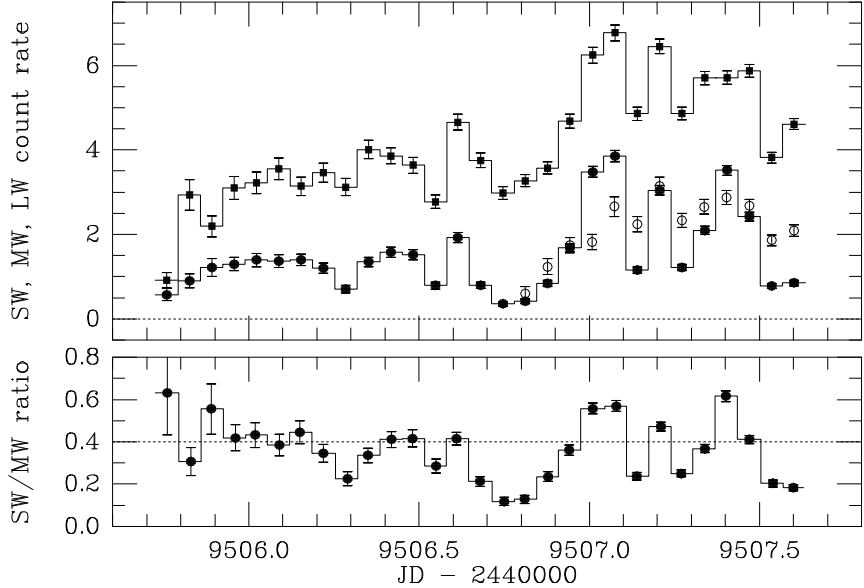}}
\caption{Upper panel: SW (100--170 \AA : {\it filled circles\/}), MW (170--300
\AA : {\it filled squares\/}), and LW (300--400 \AA : {\it open circles}) count
rate light curves of VW Hyi. Lower panel: SW over MW (hard over soft) count rate
ratio.}
\end{figure}

Blithely ignoring these indications that the EUV spectrum of VW~Hyi varied both
in flux and shape during the observations, we constructed a net spectrum for
the observation using data from each of the spectrometers. The result is shown
in Figure~2. The SW, MW, and LW portions of the spectrum extend from 80--180,
160--360, and 300--420~\AA , respectively, and are  binned to 1, 2, and
4~\AA , respectively; approximately twice the FWHM of the resolution of each
spectrometer. There is residual signal in the LW spectrometer longward even of
420~\AA , but this signal is most likely due to second-order flux; this effect
is being investigated. The errors in the flux density due solely to counting
statistics are everywhere less than $\rm 1\times 10^{-12}~erg~cm^{-2}~s^{-1} 
~\AA ^{-1}$, and are typically less than half this amount.

\begin{figure}
\centerline{\includegraphics{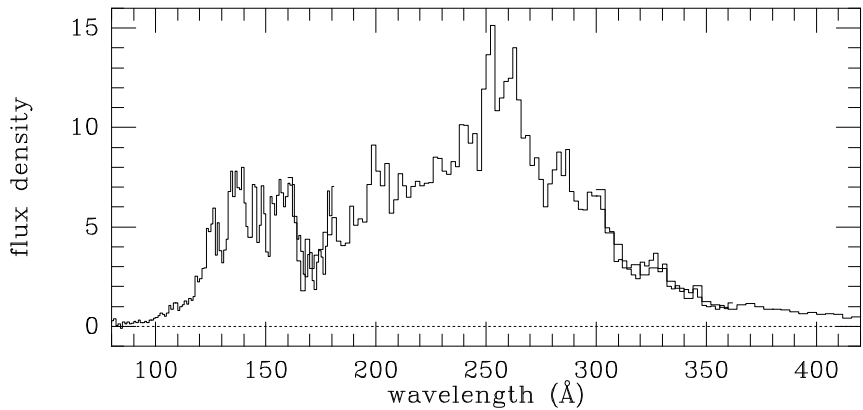}}
\caption{Net SW (80--180 \AA ), MW (160--360 \AA ), and LW (300--420 \AA )
spectra of VW Hyi. Units of flux density are
$\rm 10^{-12}~erg~cm^{-2}~s^{-1}~\AA ^{-1}$.}
\end{figure}

\vskip -12pt
\section{Discussion}

After years of speculation about its nature (e.g., van der Woerd, Heise, \&
Bateson 1986; Pringle et~al.\ 1987; Mauche et~al.\ 1991; van Teeseling, Verbunt,
\& Heise 1993), it is gratifying finally to see the EUV spectrum of VW~Hyi.
Many aspects of the spectrum are consistent with previous measurements. The
fact that the spectrum peaks longward of 170~\AA \ is consistent with the fact
that in outburst the {\it EXOSAT\/} Al-Par filter count rate is higher than
the 3000~Lex or 4000~Lex count rates. That the spectrum extends to yet longer
wavelengths is consistent with the low neutral hydrogen column density ($\approx
6\times 10^{17}~\rm cm^{-2}$) measured by Polidan, Mauche, \& Wade (1990).
That the spectrum does not extend shortward of $\approx 100$~\AA \ ($\approx
0.12$~keV) is consistent with the nondetection in the 0.18--0.43~keV bandpass
by {\it HEAO-1\/} (C\'ordova et~al.\ 1980) and with the {\it decrease\/} during
outburst of the flux in the 0.1--2.4~keV bandpass measured by {\it ROSAT\/}
(Wheatley et~al.\ 1996). Inconsistent is the brightness of the spectrum.
The observed 80--420~\AA \ flux is $1.5\times 10^{-9}~\rm erg~cm^{-2}~s^{-1}$,
implying a luminosity of $7.7\times 10^{32}\, (d/65~{\rm pc})^2~\rm erg~s^{-1}$.
For comparison, the luminosity of the accretion disk is $\approx 20$ times
higher (Mauche et~al.\ 1991). The predicted {\it EXOSAT\/} Al-Par and 3000~Lex
filter count rates are 28.3 and 15.5  counts~$\rm s^{-1}$, respectively. The
highest measured {\it EXOSAT\/} Al-Par and 3000~Lex filter count rates appear
to be 4.3 and 1.2 counts~$\rm s^{-1}$, respectively (van der Woerd \& Heise
1987). Given the low net exposure on source, it is fortunate that VW~Hyi was
so bright during our observations.

It is fortunate also that the neutral hydrogen column density to VW~Hyi is
so low. Columns typical of other nearby CVs ($N_{\rm H}\gax {\rm few}\times
10^{19}~\rm cm^{-2}$; Mauche, Raymond, \& C\'ordova 1988), would render VW~Hyi 
nearly unobservable: at $N_{\rm H}= 3\times 10^{19}~\rm cm^{-2}$, the optical
depth of the ISM is 1.1 at 100~\AA , 3.2 at 150~\AA , 6.5 at 200~\AA . SS~Cyg
(Mauche, Raymond, \& Mattei 1995) and U~Gem (Long et~al.\ 1995) both have
higher columns, but are also intrinsically harder. If other, even nearby, CVs
are as soft as VW~Hyi, it will be hard to impossible to detect them in the
EUV ($\lambda = 100$--912~\AA ).

As is the case with SS~Cyg and U~Gem, the EUV spectrum of VW~Hyi belies simple
interpretation. Consideration of the following options is instructive of the
problems facing us.

(i) \underbar{A blackbody spectrum.}
To produce so little flux shortward of $\approx 100$~\AA , the temperature of a
blackbody must be $T\lax 1\times 10^{5}$~K ($kT\lax 10$~eV). To extinguish the
flux longward of $\approx 350$~\AA , the neutral hydrogen column density then
must be $N_{\rm H}\gax 3\times 10^{18}~\rm cm^{-2}$, but such a model fails
to reproduce the overall spectral distribution.

(ii) \underbar{A modified blackbody spectrum.}
The ionization edges of C~III--IV, N~III--V, O~III--IV, Ne~II--IV, Mg~III--IV,
Si~IV,  S~IV--VI, and Fe~IV--VI all lie in this bandpass, but are not apparent
in the spectrum. This result probably does not constrain irradiation of the
white dwarf by a hard continuum spectrum, since such models produce edges at
wavelengths below 100~\AA \ (van Teeseling, Heise, \& Paerels 1994). 

(iii) \underbar{The spectrum of an optically thin plasma.}
At $T\approx 1$--${\rm few}\times 10^{5}$~K, the He~II Lyman lines ($\lambda =
304, 256, 243, \ldots $~\AA ) and bound-free continuum ($\lambda < 228$~\AA )
are in emission, contrary to observations. At higher temperatures, particularly
$T\gax 5\times 10^{5}$~K, the free-free continuum is weak, but strong emission
lines begin to appear shortward of 100~\AA . 

Other processes likely affect the spectrum. The ISM and possibly the wind of
VW~Hyi will imprint a He~II absorption edge at 228~\AA \ onto the intrinsic
spectrum. Measurement of, or upper limits on, such an edge will provide an
estimate of the integrated column density of this ion. He~I has autoionization
resonances at $\lambda = 206, 195, 192, \ldots $~\AA \ (Rumph, Bowyer, \& Vennes
1994). The first such resonance may be responsible for the emission/absorption
feature at 206~\AA .

\smallskip
\noindent
{\bf Acknowledgements.} The author is pleased to acknowledge the contributions
to this research by the members, staff, and director, J.\ Mattei, of the AAVSO.
This work was performed under the auspices of the U.S.\ Department of Energy by
Lawrence Livermore Nat'l Laboratory under contract No.\ W-7405-Eng-48.

\end{article}
\end{document}